# Electric-field-induced magnetization changes in $Co/Al_2O_3$ granular multilayers

Ajeesh M. Sahadevan[1], Alan Kalitsov[2], Gopinadhan Kalon[1], Charanjit S. Bhatia[1], Julian Velev[2], and Hyunsoo Yang[1,†]

*[1]Department of Electrical and Computer Engineering, National University of Singapore, 4 Engineering Drive 3, Singapore 117576, Singapore*
*[2]Department of Physics, University of Puerto Rico, San Juan, PR 00931, USA*

We study experimentally the effect of electric field on the magnetization of $Co/Al_2O_3$ granular multilayers. We observe two distinct regimes: (a) low-field regime when the net magnetization of the system changes in a reversible way with the applied electric field and (b) high-field regime when the magnetization decreases irreversibly. The former is attributed to the changes in the relative 3*d*-orbital occupation of the minority and majority bands in the Co granules. A theoretical model has been developed to explain the electric field induced changes in the band structure of the granular system and hence the magnetic moment. The latter result may be understood assuming the electric field induces oxygen migration from $Al_2O_3$ to the Co granules, since an increase in oxidation state of the Co granules is shown, through *ab-initio* calculations, to give rise to a reduced magnetization of the system.

Electric field control of magnetism has aroused significant interest for future electronics as well as energy efficient magnetic data storage[1]. The effect of electric field on several electronic systems based on magnetic thin films has been demonstrated in the past few years. Application of an external electric field in perpendicular magnetic anisotropy based CoFeB/MgO/CoFeB magnetic tunnel junction aids in lowering the current densities for spin transfer torque switching[2] and for an ultra thin Fe film, the anisotropy direction can be controlled depending on the polarity of the electric field applied across the film.[3] Several theoretical reports also suggest the possibility of changes in the spin density of states at Fe/MgO interfaces in the presence of electric field that can affect both the anisotropy and the net moment of the Fe film.[4, 5]

Granular magnetic films provide interesting routes for novel physics and device applications by tailoring both the individual and collective properties of the nano-magnets.[6, 7] For example, incorporation of magnetic nanoparticles in the barrier of magnetic tunnel junctions has been theoretically predicted to demonstrate a very high magnetoresistance[8] and higher order tunneling effects have also been experimentally observed in similar systems.[9, 10] In terms of applications, resistive switching is an interesting phenomenon also observed in magnetic granular films.[11, 12] However, along with the changes in the resistance of the granular system in the presence of an electric field, the magnetic moment of the granules will also be affected and this has not been investigated either through experiments or theoretical modeling. Another effect of the high electric field can be the migration of the oxygen atoms from the oxide matrix into the Co granules which would affect the magnetization of the granules.

In this work, we investigate the effect of electric field on the magnetization of Co granules in the sputter deposited Co/$Al_2O_3$ based magnetic granular system. In the low field regime we show systematic and reproducible changes in the net magnetic moment of the system using *in-situ* electric fields in a superconducting quantum interference device (SQUID). Using a theoretical model we explain the experimental results based on the relative changes in the occupation of the majority and minority states of the Co granules. In the high electric field regime we observe that the magnetization of the system decreases irreversibly with the applied field, which is attributed to oxidation of the Co granules. We perform first-principles calculations based on the density-functional theory (DFT) to understand the effect of oxidation on the net magnetic moment.

The $Co/Al_2O_3$ magnetic granular system is fabricated in an ultra-high vacuum ($10^{-9}$ Torr) magnetron sputtering chamber using dc and rf sources for Co and $Al_2O_3$, respectively. Ten layers of [$Al_2O_3$ (4 nm)/Co (0.5 nm)] are deposited on Si (100) substrates with a 400 nm thick thermally oxidized $SiO_2$ layer and the structure is subsequently capped by a 4 nm $Al_2O_3$ layer. Figure 1(a) shows the cross sectional transmission electron microscope (TEM) image of $Co/Al_2O_3$ granular multilayers. Co nanoparticles are clearly visible, embedded in an $Al_2O_3$ insulating matrix. The inset of Fig. 1(a) shows a schematic of the system used in this study. Electrical contact pads are formed by thermally evaporated Cr (5 nm)/Au (100 nm) and an electric field is applied between the pads. The sample size is ~ 5 mm × 5 mm. Figure 1(b) is the TEM image of a region of the sample where electric field has been applied. Magnetic moment measurements have been carried out using SQUID.

In order to investigate the magnetization changes in the presence of small electric field, we carry out SQUID measurements with an in-situ bias voltage across the sample at room temperature. Application of an *in-situ* electric field in a SQUID system enables direct measurement of the voltage dependence of magnetic moments and this strategy has been used in other systems as well.[13] The magnetization of the $Co/Al_2O_3$ granular system is found to reduce systematically and reproducibly, as the electrical field in-plane to the surface gradually increases. Figure 2(a) and (b) show results for two samples, where the value of magnetic moment is averaged over a period of 200 s with a constant external field of 1 T. The measurement is repeated four times with increasing and decreasing electric fields (1st and 2nd sequence) as well as with negative values of electric fields (3rd and 4th sequence). For sample #1 the maximum electric field is around 0.4 mV/nm, while for sample #2 it is around 0.1 mV/nm. In order to confirm reliability of the measurements, a similar measurement is carried for a sample with 20 nm $Al_2O_3$ and no voltage dependence is observed. The changes we observe are small (~2%), however comparable to changes predicted in thin Fe films for a similar applied level of electric field.[4] Nevertheless, our experiment provides a direct evidence to demonstrate that electric fields can change the magnetization of Co granules in an $Al_2O_3$ matrix. In this case oxygen migration is not primarily involved due to a smaller applied electric field,[14, 15] therefore, the change in the magnetization is reversible with the bias as shown in Fig. 2.

In order to gain insight in the effect of electric field on the magnetic moment of the Co granules, we study the density of states (DOS) of the granular system in the presence of an external field. We use a coarse-grained tight-binding model of a bulk granular system where a magnetic granule is represented by a single site. Conduction electrons can hop between the nearest-neighbor granules. Despite its simplicity this model has been shown to be capable of successfully describing resistive switching phenomenon in Co/$Al_2O_3$ magnetic granular multilayers.[11] Conduction electrons are coupled with the magnetic moments of granules through the local exchange interaction $J_0$. The Hamiltonian of the granular system has the form

$$H = -J_0 \sum_i M_i (c_i^{\dagger\uparrow} c_i^{\uparrow} - c_i^{\dagger\downarrow} c_i^{\downarrow}) + t \sum_{i,j,\sigma} c_i^{\dagger\sigma} c_j^{\sigma} \quad (1)$$

where $t$ is the spin-independent effective hopping integral between the nearest-neighbor granules and $M_i$ is magnetic moment of *i*-th granule in the direction of applied magnetic field. The total magnetization of the granular alloy consists of two contributions. The first one is the magnetization of the localized moments $M_{loc} = \sum_i M_i$. This magnetization does not depend on applied electric field. The second contribution comes from the conduction electrons $M = M_{\max}\left(\left\langle N^{\uparrow}\right\rangle - \left\langle N^{\downarrow}\right\rangle\right)$, where $M_{\max}$ is the maximum value of the contribution from the conduction electrons. This contribution depends on the spin-dependent occupation numbers $\left\langle N^{\sigma}\right\rangle$ which can be controlled by applying the electric field. Our calculations are based on the Green function (GF) method. First we find the GF $G_{ij}^{\sigma}$ corresponding to the above Hamiltonian. The spin-dependent DOS is given by the imaginary part of the diagonal elements of the GF $\rho^{\sigma}(E) = -\frac{1}{\pi} G_{ii}^{\sigma}(E)$. Finally, the spin-dependent occupation numbers are given by the integral over energy $\left\langle N^{\sigma}\right\rangle = \int_{-\infty}^{\infty} f(E)\,\rho^{\sigma}(E)dE$, where $f = [1+\exp(E-\mu)/kT]^{-1}$ is the Fermi-Dirac distribution function. We assume that the chemical potential in the granules depends linearly on the applied voltage $\mu = \alpha eV$, where we have chosen the coefficient $\alpha = 0.01$, because most of the voltage drop is in the insulator. We vary the parameter $J_0/t$. Since the conduction electrons hope between the granules through an insulating $Al_2O_3$ barrier, the hopping integral is also small $J_0/t >> 1$.

The DOS of $Co/Al_2O_3$ granular system is shown in Fig. 3(a). The splitting between the centers of the majority and minority bands is given by $2J_0$, while the half of the bandwidth is equal to $6t$ for the simple cubic tight-binding model. Thus when $J_0/t = 6$, the bottom of the minority band is aligned with the top of the majority band. Decrease of $J_0/t$ leads to the overlap between the majority and minority bands. In contrast, increase of $J_0/t$ opens the band gap in the granular system. The Fermi energy is at zero eV in the absence of the electric field. Thus the majority band is occupied while the minority band is empty.

We plot the corresponding voltage induced change of the magnetization in Fig. 3(b). Positive (negative) electric field shifts the chemical potential of the system towards smaller (higher) values of energies, yielding decrease (increase) of the majority (minority) band occupation. As a result the magnetization of the granular system decreases symmetrically with respect to the sign of applied voltage. Note, if $J_0 / t > 6$, a very small electric field does not change the magnetization of the system until the Fermi energy is within the band gap. The calculated change of the contribution to the magnetization from conduction electrons can be almost 10% [Fig. 3(b)]. Experimentally one cannot separate the contributions from conduction and localized electrons. Since the contribution from localized electrons does not depend on applied electric field, the electric field induced change of the total magnetization is smaller, about 2%.

For device applications the reversible approach based on electric field control of the 3*d* states is more relevant. Addition of O atoms into magnetic granules is another approach to change the occupancy of the 3*d* states and hence magnetic moment, however, it is an irreversible and uncontrollable process. We also analyze the oxidation state of Co granules before and after the application of a very high electric field (~ 4 mV/nm) using X-ray photoelectron spectroscopy (XPS). Figure 4(a) shows the depth profile of as deposited Co-$Al_2O_3$ multilayers showing clear alternate oscillations of $Co_{2p}$ and $O_{1s}$ peaks. Figure 4(b) and (c) show the $O_{1s}$ spectrum for the $Al_2O_3$/Co multilayers at a depth corresponding to the first Co layer from the top surface, before and after electric field application, respectively. $O_{1s}$ peaks have been used for the calculation of relative proportions of different species in various materials and systems.[16-19] The experimental spectra is corrected by Shirley background and then fitted using a Voigt function.[18, 20] The $O_{1s}$ peak for Al-O bonding is at 531.6 eV, while the Co-O bonding peak is 530.4 eV for the fits.[21]

However, as shown in Fig 4(b) and (c), the $O_{1s}$ spectra for both the cases (as deposited and after electric field application) can be fitted very well using a single Voigt function peak at 531.3-531.6 eV. The fluctuation can be accounted by the energy resolution of 0.2 eV for the XPS spectra in the current measurement. The absence of any Co oxide signal is possibly a result of either a very small amount of granule oxidation or the relatively large signal from the $Al_2O_3$ background. The magnetic moment of the granules, however reduces in magnitude (by 25%) as shown in Fig. 4(d). The amount of the reduction varies from sample to sample and cannot be quantified due to the large number of granules between the two contact pads. We attribute this to high electric field induced oxygen migration from the oxide matrix to the Co granules, even though the proposed oxidation is beyond the sensitivity of XPS. For very high electric fields in the range of a few mV/nm, the changes in magnetization as well as the oxidation state of Co are irreversible.

In order to understand the effect of oxidation (and correspondingly of high electric filed) we perform first-principles calculations based on DFT. We use the projector augmented wave (PAW) method implemented in the Vienna *ab-initio* simulation package (VASP)[22] with the Perdew-Burke-Ernzerhof (PBE) exchange-correlation functional. As a model for Co granule surface we consider a 5×5 supercell of the fcc Co (111) surface. In the direction perpendicular to the surface we use 5 layers of Co and 10 Å of vacuum to separate the two surfaces. The magnetic moment of Co atoms on the surface is about 1.78 $\mu_B$, slightly larger than the moment in the bulk 1.69 $\mu_B$. From previous calculations we know that the most energetically favorable O adsorption site is the 3-fold site in which the O atom is between 3 Co atoms about 1.04 Å above the surface.[23,24]

Full oxidation requires one O atom for each Co in which case the Co atoms are fully saturated (three Co-O bonds). As a result of the oxidation the magnetism on the surface Co layer is severely suppressed and the magnetic moment drops to 0.27 $\mu_B$ per Co. To study the intermediate regime we perform a number of calculations in the low oxidations state (1-3 O atoms in different configurations) and in the high oxidation state (22-24 O atoms). In these calculations we use 400 eV plane wave expansion cutoff and a 3×3×1 Monkhorst-Pack grid for the k-points. These calculations do not include the effects of the curvature, structural defects, and the relaxation which would be inevitably present in the experimental setup.

We calculate the bonding energy and the magnetization change in all the configurations. The bonding energy is calculated as $\Delta E = E_{Co-O} - E_{Co} - n_O E_{O2} / 2$ where the first term is the energy of the oxidized slab, the second is the energy of the clean Co slab, and the last is the energy of the O adatoms. Similarly the magnetization change is calculated as $\Delta M = M_{Co-O} - M_{Co}$. We notice that the O atom adsorption affects essentially only the Co atoms to which it is directly bonded. Moreover, the Co atoms with 1, 2, and 3 bonds have distinctly different magnetic moments. We, therefore, perform a least square regression of the bonding energy and the magnetic moment change against the number of Co atoms with different degree of oxidation. The results of the regressions are $\Delta E = -0.695 n_{Co1} - 1.261 n_{Co2} - 1.653 n_{Co3}$ and $\Delta M = -0.140 n_{Co1} - 0.647 n_{Co2} - 1.436 n_{Co3}$, where $n_{Co1}$, $n_{Co2}$, and $n_{Co3}$ are the numbers of Co atoms with 1, 2, and 3 O bonds, respectively. First we notice that all coefficients are negative. For the bonding energy this means that oxidation is generally favorable with saturating the Co bonds being most energetically favorable. At the same time higher oxidation state is associated with reduction of the magnetic moment at the surface where each partially saturated Co atom reduces the magnetization by 0.140 $\mu_B$ (1 bond), 0.647 $\mu_B$ (2 bonds), and each fully saturated Co atom reduces the magnetization by 1.436 $\mu_B$.

Thus the increase in the oxidation of the Co granule surface is energetically favorable and leads to a monotonous decrease in the sample magnetic moment. However, the size of the change of the magnetic moment is not constant and it would depend on the sample's initial oxidation state. At low Co oxidation most Co atoms will have unsaturated bonds and the magnetization would drop relatively slowly. At high oxidation states additional oxidation would produce more saturated Co bonds and decrease the magnetic moment faster. This behavior is consistent with the observed experimental results where the magnitude of reduction in magnetization varies from sample to sample.

In conclusion, we have shown the effect of electric field on the magnetic moment of Co granules in an $Al_2O_3$ matrix. The magnetic moment in the presence of a small in-situ electric field (~ 0.1-0.4 mV/nm) is systematically reduced as the field increases. The changes in the magnetization of the Co granules are attributed to changes in the position of the Fermi level relative to the minority and majority bands. Higher electric field (~ 4 mV/nm) decreases the magnetization of the granules in an irreversible way. The magnetic moment decrease is attributed to oxidation of Co granules which is consistent with our first-principles calculations. Our work opens up the possibility to control the magnetization by electrical fields in magnetic granular systems.

This work was partially supported by the Singapore NRF under CRP Award No. NRF-CRP 4-2008-06, NSF (Grant No EPS-1002410 and EPS-1010094) and DOE (Grant No DE-FG02-08ER46526).

†eleyang@nus.edu.sg

**Figure captions**

**FIG. 1.** (a) Cross sectional TEM image of the Co/$Al_2O_3$ multilayer system (as deposited). The dark spots are Co islands and the lighter region is the $Al_2O_3$ insulating matrix. A schematic representation of the multilayer system is shown in the inset. (b) TEM of the multilayers after applying a high electric field along the plane of the film.

**FIG. 2.** Magnetic moment versus applied small electric field for two samples at room temperature. The magnetic moment gradually reduces as the electric field increases, and the changes are reproducible ($1^{st}$ to $4^{th}$ steps indicate the sequence of measurements). The contact pads are indicated in the insets.

**FIG. 3.** (a) DOS of Co/$Al_2O_3$ granular multilayers. The upper (lower) panels show the majority (minority) states. (b) Electric field induced change in the conduction electron's contribution to the magnetization. The magnetic moment of the granules reduces with both polarities of voltage bias depending on the position of the Fermi level relative to the majority and minority electron bands.

**FIG. 4.** (a) XPS depth profile of the multilayer system showing alternating oscillation peaks of $Co_{2p}$ and $O_{1s}$. $O_{1s}$ spectra of the layers at the first Co layer from as deposited sample (b) and for region after application of electric field (c). The spectra is fitted using Voigt function. (d) M-H loops using SQUID showing the changes in magnetization before and after application of high electric field. Net magnetic moment for the sample decreases after bias application.

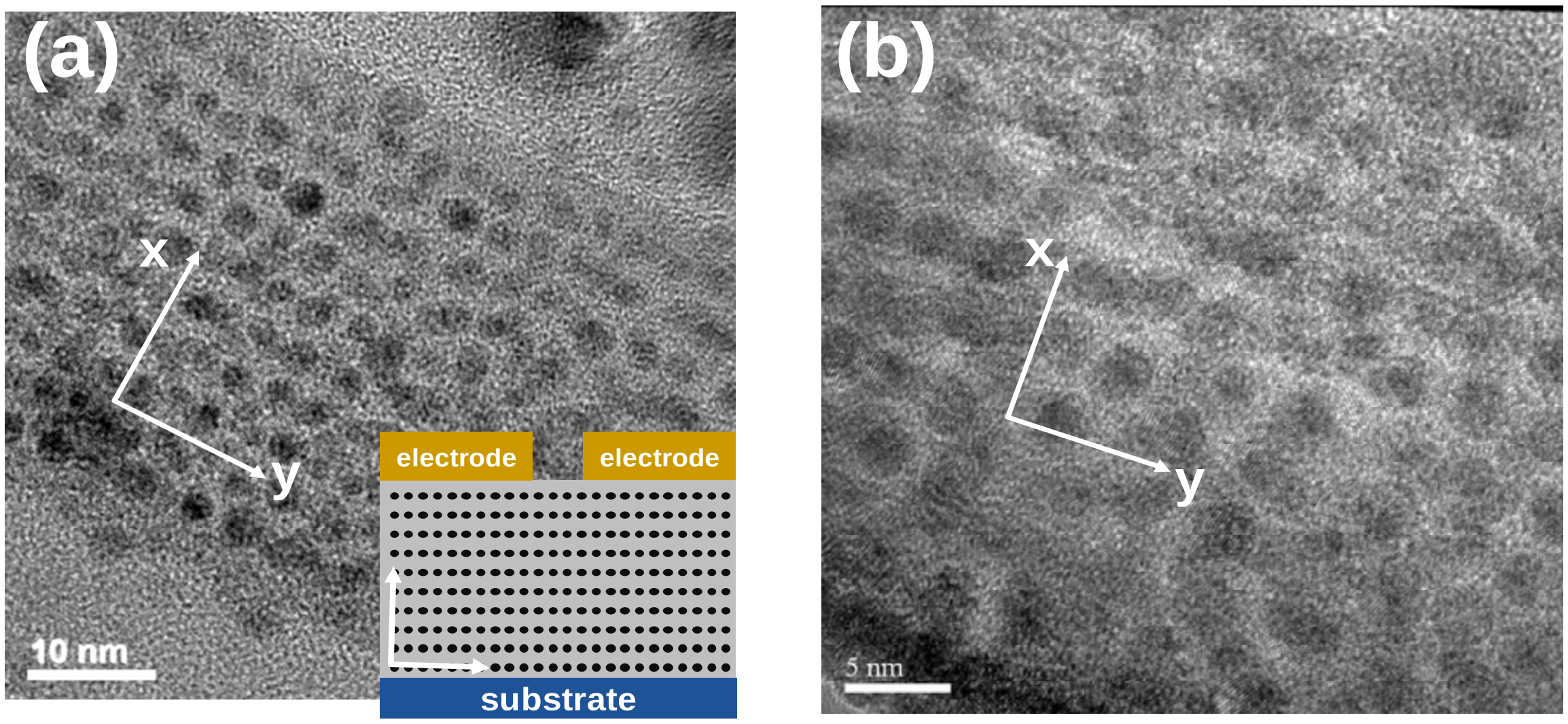

Figure 1

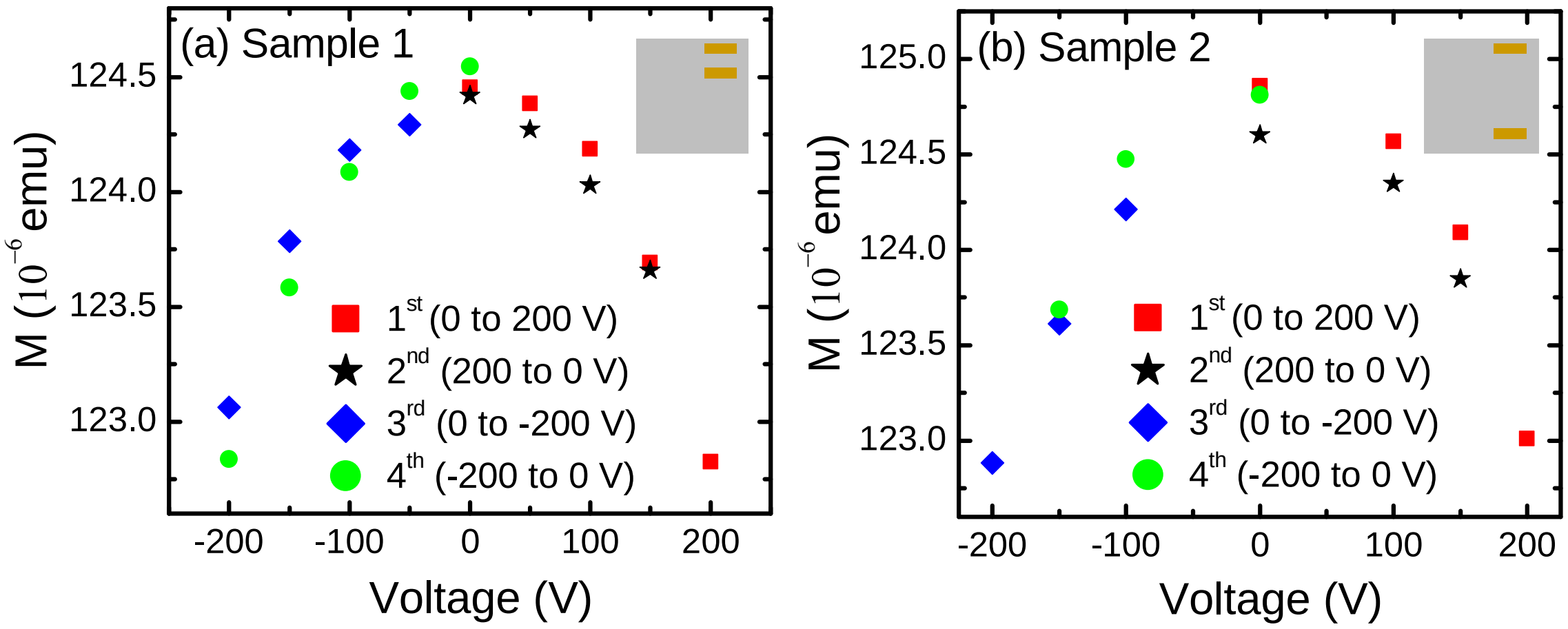

Figure 2

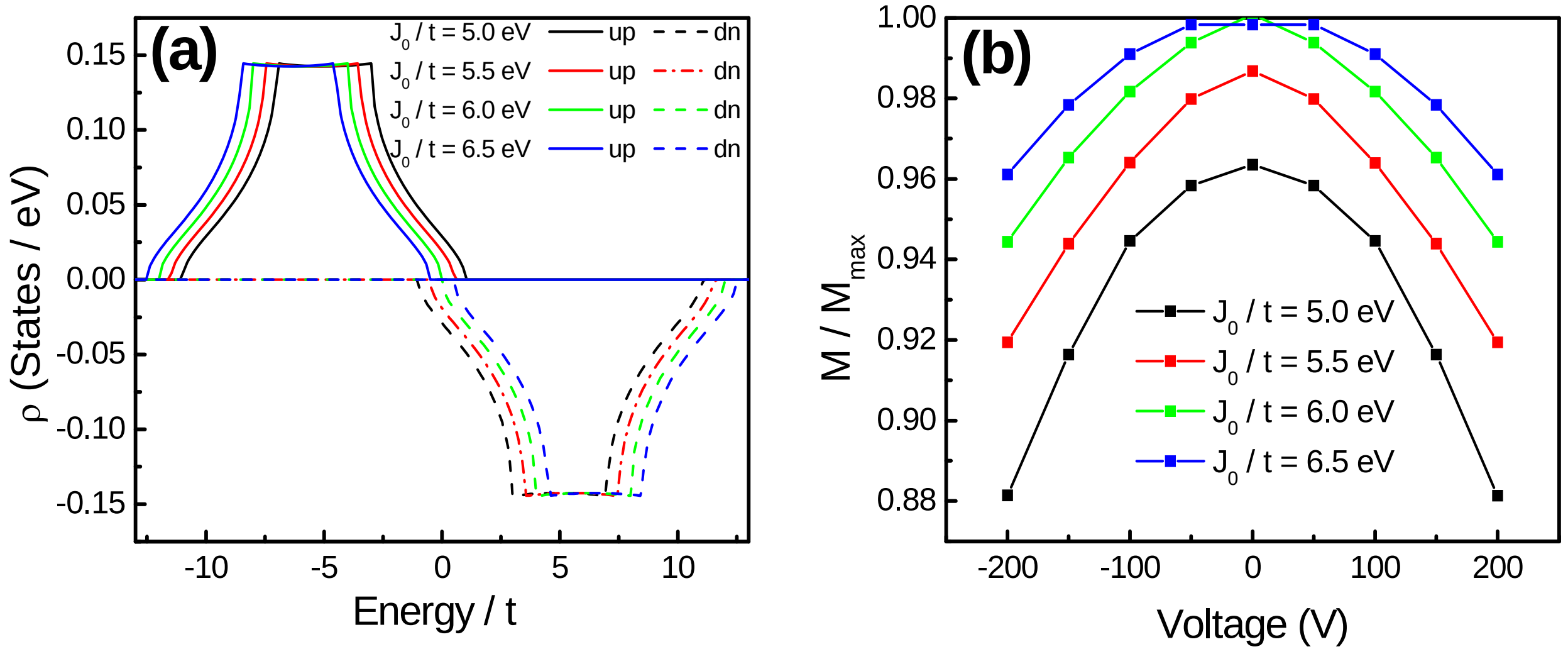
(a)
J0 / t = 5.0 eV up dn
J0 / t = 5.5 eV up dn
J0 / t = 6.0 eV up dn
J0 / t = 6.5 eV up dn
ρ (States / eV)
Energy / t
(b)
M / Mmax
J0 / t = 5.0 eV
J0 / t = 5.5 eV
J0 / t = 6.0 eV
J0 / t = 6.5 eV
Voltage (V)

Figure 3

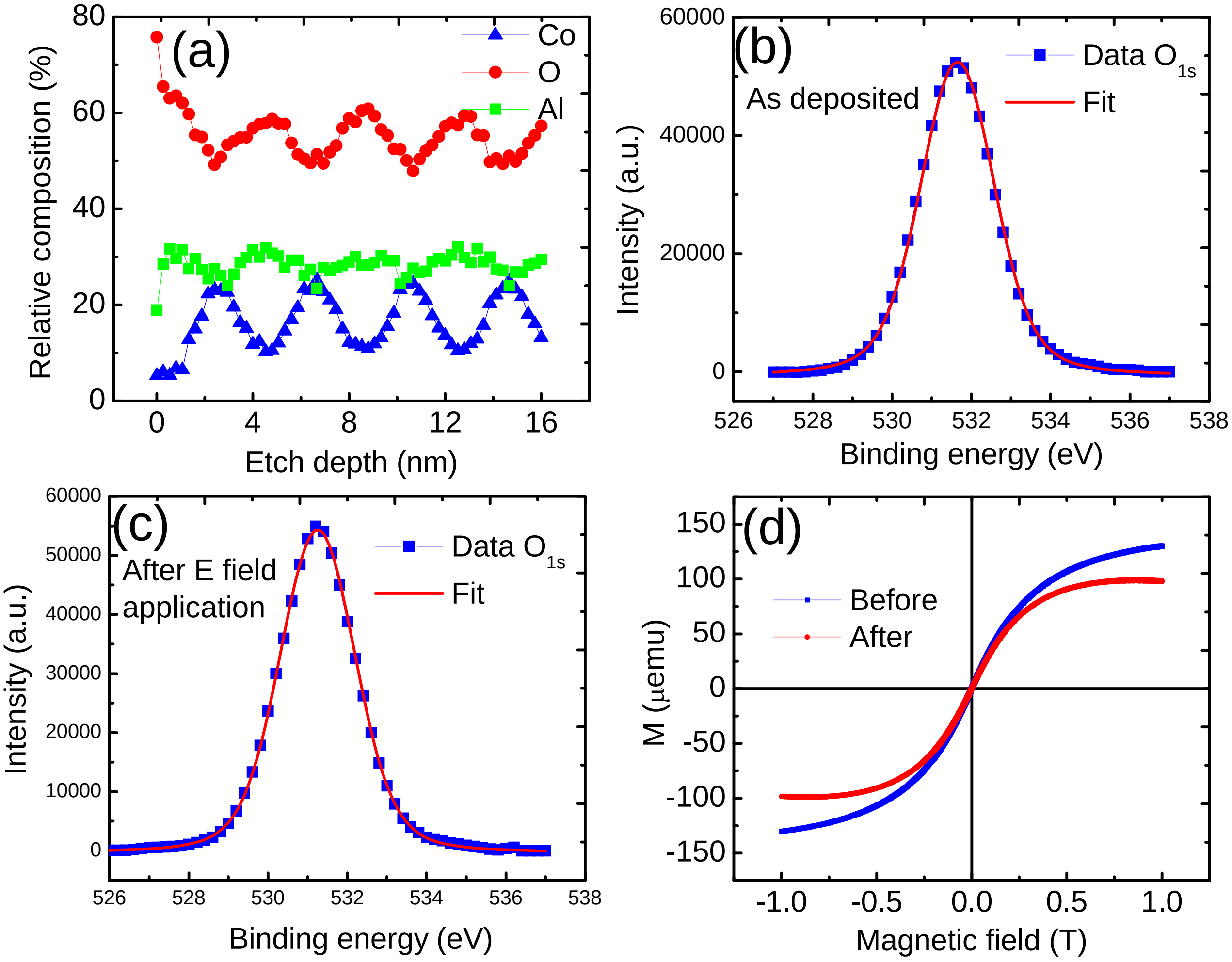

Figure 4